\documentclass[prd,twocolumn,amsmath,amssymb]{revtex4}

\usepackage{amsbsy}
\usepackage{graphicx}
\usepackage{graphics}
\usepackage{latexsym}

\begin{document}
%

%
%
%
%
%
%
%

\newcommand{\sect}[1]{\setcounter{equation}{0}\section{#1}}
\renewcommand{\theequation}{\thesection.\arabic{equation}}
\setcounter{equation}{0}

\newcommand{\di}[2]{\frac{d {#1}}{d {#2}}}
\newcommand{\de}[2]{\frac{\partial {#1}}{\partial {#2}}}

\def\spazio#1{\vrule height#1em width0em depth#1em}
\def\ad{a^\dagger}
\def\bra#1{\langle#1|}
\def\ket#1{|#1\rangle}
\def\scal#1#2{\langle#1|#2\rangle}
\def\matrel#1#2#3{\langle#1|#2|#3\rangle}
\def\pert#1{|\psi_n^{(#1)}\rangle}
\def\scalpert#1#2{\langle#1|\psi_n^{(#2)}\rangle}

\def\spazio#1{\vrule height#1em width0em depth#1em}
\def\enr{E_{{}_{\mathrm{NR}}}}

%
%
%

\begin{abstract}
The spectral problem of the Dirac equation in an external quadratic vector potential is considered using the
methods of the perturbation theory. The problem is singular and the perturbation series is asymptotic, so that
the methods for dealing with divergent series must be used. Among these, the Distributional Borel Sum
appears to be the most well suited tool to give answers and to describe the spectral properties of the system.
A detailed investigation is made in one and in three space dimensions with a central potential. We present numerical results  for the Dirac equation in one space dimension: these are obtained by determining the perturbation expansion and using the Pad\'e approximants for calculating the distributional Borel transform.
A complete agreement is found with previous non-perturbative results obtained by the numerical solution of the singular boundary value problem and the determination of the density of the states from the continuous spectrum.

\end{abstract}

\bigskip
\bigskip

\title{\bf Perturbation Theory for Metastable States of the Dirac Equation with Quadratic Vector Interaction.}

\author{Riccardo Giachetti}
\affiliation{Dipartimento di Fisica, Universit\`a di
Firenze, Italy}
\affiliation{Istituto Nazionale di Fisica Nucleare, Sezione di Firenze}
\email{giachetti @ fi.infn.it}
\author{Vincenzo Grecchi}
\affiliation{Dipartimento di Matematica, Universit\`a di
Bologna, Italy}
\email{grecchi @ dm.unibo.it}

\maketitle

\sect{Introduction.}
\label{intro}
\bigskip

It has been argued since a long time \cite{Plesset} that the Dirac equation with an unbounded potential in vector coupling
has no discrete but only  completely continuous spectrum. It was shown in \cite{Tit0,Hart} and  subsequently confirmed in
literature, that precise conditions have to be imposed to the potential for this to be true.  
In two later papers, \cite{Tit1,Tit2}, Titchmarsh  proved that the Dirac equation with a  linear vector potential satisfies the needed requirements and he studied  the relativistic quantum mechanics of an electron in a constant (or piecewise constant)
electric field.  The first 
order of the asymptotic perturbation expansion was explicitly calculated due to  the possibility of integrating the Dirac
equation in one space dimension  by means of hypergeometric functions, but hard technical difficulties prevented further
analytical developments, as well as the extension to a higher power law for the interaction or
to more general inhomogeneous electric fields.
The Dirac equation with a vector potential that grows sufficiently fast at infinity is an incomplete dynamical problem.
In classical terms, the particle can arrive at infinity in a finite time; in
the operator language, we have (2,2) deficiency indices and the appropriate 
asymptotic boundary conditions must be defined. 
There is an infinite number of such boundary conditions of the self-adjoint type not very meaningful from a physical point of view.
The most physical condition is the absence of sources at infinity, corresponding to the Gamow-Siegert conditions, that assume 
outgoing wave functions only:
in this case, however, the Hamiltonian is not self-adjoint, the spectrum has complex eigenvalues, the dynamics is dissipative and
the eigenfunctions have the meaning of metastable states of the dissipative dynamics itself:
\begin{eqnarray}
\!\!\!\!\! \|U_t\psi_n\|^2=\|e^{-iE_nt/\hbar}\,\psi_n\|^2=e^{-\Gamma_nt}\,\|\psi_n\|^2
\label{normdecay}
\end{eqnarray} 
where $\Gamma_n=-2\,{\mathrm{Im}}(E_n/\hbar)>0$,  is the inverse of the mean lifetime of the metastable state $\,\psi_n$.
Actually such states are stationary, if we neglect the decrease of the norm, so that they should be called metastationary;
we prefer not to call them resonances, since they are indeed different from the usual resonances and have a direct physical meaning.
 
Thinking of the transition from the non-relativistic 
to the relativistic system as a perturbation process in terms of the small parameter $(1/c)$, the disappearance of the 
Schr\"odinger bound states connected with the confining potential defines a singular perturbation problem. 
A numerical non-perturbative investigation of the one dimensional Dirac equation with linear and quadratic potential has recently appeared
\cite{GS}: the purpose of that paper was to follow very closely the change of the spectrum when passing to the relativistic regime
and to describe the spectral concentration \cite{K,RS} at finite values of $(1/c)$, or, in a more physical language,
to determine the density of the states of the relativistic system. It was indeed produced a numerical evidence that the 
spectrum is completely continuous and that it is given by a sum of Breit-Wigner lines  reducing to  $\delta$-functions centered at the non-relativistic eigenvalues for smaller and smaller values of the ratio
of the interaction to the rest energy, so that
the spectral measure becomes atomic as it should. In \cite{GS} the pair production rate was also calculated from the line width 
of the lowest state finding a perfect agreement, for the linear potential and in the range of validity of the first perturbation order,
with the results obtained from the imaginary part of the Schwinger effective action, \cite{Sch}. It
should also be noticed that the QED results on the pair production for fermions in non constant electric fields are not yet so
sound, although interesting proposals can be found in literature, (see \cite{Kim} for an up to date review). We can finally remark that
relativistic models with power-law potentials, in three space dimensions and in spherical geometry, have been used in the study of composite systems like Quarkonium, in order to determine the mass spectrum of some meson families, \cite{Martin}.

In this paper we come back to the spectral properties of the Dirac equation with a quadratic vector potential
in a perturbation framework, completely different from that adopted in \cite{GS}. Indeed, since the quantitative and numerical results on this
subject are rather new, we find it interesting to make a comparison of independent computational approaches and eventually to have a confirmation of the results. 
In particular, we use here the method of the \textit{Distributional Borel Sum} (hereafter DBS), that has proved a very useful tool for dealing with physical systems of singular nature. Originating from a suggestion given by 't Hooft for double well problems, \cite{tH},  the DBS was studied in a series of papers \cite{CGM86}-\cite{CGM96} and it was successfully applied to $\lambda\Phi^4$ lattice fields with large coupling in \cite{CGM86}, to the non-relativistic Stark resonances in \cite{CGM93} and to the double well Schr\"odinger operators in \cite{CGM88,CGM96}. 
To our knowledge however, despite the very deep mathematical development of the subject, no explicit numerical calculations with the DBS have been made up to date and no applications of it to the  Dirac equation have been considered. In this sense this 
paper represents a novelty and a test of the method for its possible practical uses. 
We will also show that the basic results of \cite{CGM93} can be brought to bear
to our present context, although some peculiarities due to the relativistic nature of the problem will emerge and must be taken into account. 

The plan of the paper is the following. In the next section, for the sake of completeness and because the knowledge of the DBS is not so widely diffused,
we give a brief  description of the method  and we state more properly the spectral problem.  In section \ref{dbs} we study the general conditions
for applying the DBS to the Dirac equation and we discuss the possible strategies for the one-dimensional and the three-dimensional problems.
 Finally, in section \ref{numeri}, we present  the numerical treatment for the one-dimensional case. Using a perturbation parameter proportional to the ratio of
the interaction to the rest mass energy, we calculate  the perturbation series of the fundamental state energy up to a large order and
we then construct its Borel transform determining its asymptotic behavior.  We then approximate the Borel transform by Pad\'e approximants. Although, in a strict
mathematical sense, this is not a completely rigorous procedure since it is not proved that the Borel transform is a Stieltjes function, however an idea of the accuracy of the approximation can be obtained by comparing the position of the poles of the Pad\'e approximants with the location suggested by the asymptotic form of the perturbation series: indeed, as shown in section \ref{numeri}, we find the Pad\'e poles exactly where the asymptotic behavior indicates they should be. We finally check the stability
of the poles with the increasing orders of the Pad\'e approximants, implying the stability  of the imaginary part of the perturbed energy. The latter is then calculated and it
is found a perfect agreement with the results of \cite{GS}, confirming them and  proving the correctness and the effectiveness of both methods also in explicit numerical calculations: this
is even more interesting in view of the fact that each of the two methods is more efficient in 
a different range of the values of the coupling constant, so to allow for the choice of the most appropriate one in a practical situation.

\bigskip


\sect{The spectral problem.}
\label{spectralproblem}
\bigskip

We consider the 
(3+1)-dim Dirac equation interacting by means of a central vector potential $(0,V(r))$. It is well known that the use of the spherical
spinors, \cite{LL}, easily leads to the diagonalization of the angular momentum and to the reduction of the Dirac equation to the system of the two first order differential equations
\begin{eqnarray}
 &{}&\!\!\!\!\!\!\!\!\!\frac 1c\Bigl(W+mc^2-V(r)\Bigr)X_1(r)-\hbar\Bigl(\frac{d}{dr}+\frac\kappa r\Bigr)X_2(r)=0\cr\spazio{1.8}
&{}&\!\!\!\!\!\!\!\!\!\hbar\Bigl(\frac{d}{dr}-\frac\kappa r\Bigr)X_1(r)+\frac 1c\Bigl(W-mc^2-V(r)\Bigr)X_2(r)=0\cr
&{}&
\label{DiracEquation_rad}
\end{eqnarray}
\noindent In (\ref{DiracEquation_rad})  $m$ is the mass and $W$ the energy of the particle, while  $X_1(r)=rg(r)$ and  $X_2(r)=rf(r)$,
where $f(r)$, $g(r)$ are respectively the `large' and the `small' spherical component of the spinor. Finally the parameter $\kappa$
accounts for the angular momentum and the parity: $\kappa=-(\ell+1)$ for $j=\ell+1/2$ and $\kappa=\ell$ for $j=\ell-1/2$, \cite{LL}, so that,
in three space dimensions, $\kappa$ can assume all integer values except zero.
If, instead,  we let $\kappa\!=\!0$ in equation (\ref{DiracEquation_rad}) and we change $r$ in $x$, with $-\infty<x<+\infty$, we obtain the one space-dimensional Dirac equation that has been discussed in \cite{GS}.
We now assume a quadratic potential 
\begin{eqnarray}
V(r)=(1/2)\, m\,\omega^2\,r^2,
\nonumber
\end{eqnarray}
we rescale the
system by introducing the dimensionless variables
\begin{eqnarray}
&{}& y=r\Bigl(\frac{m\omega}{\hbar}\Bigr)^{1/2}\qquad \beta=\frac{1}{2} \Bigl(\frac{\hbar\omega}{mc^2}\Bigr)^{1/2}\spazio{1.0}\cr
&{}& {}\quad\qquad{{E}}=\frac 2{\hbar\omega}(W-mc^2)\nonumber
\end{eqnarray}
\noindent and we define the unknown functions
\begin{eqnarray}
&{}&{\Phi}_1(r)=2^{-1/2}(X_1(r)+iX_2(r))\,,\spazio{1.0}\cr 
&{}&{\Phi}_2(r)=-i\,2^{-1/2}\,(X_1(r)-iX_2(r))\nonumber
\end{eqnarray}
The system (\ref{DiracEquation_rad}) becomes then
\begin{eqnarray}
 &{}&\frac d{dy}\,{\Phi}_1(y)-   i\Bigl(\frac 1{2\beta}+\beta({{E}}-y^2)\Bigr)\,{\Phi}_1(y)+\cr
&{}&{}\qquad\qquad\qquad\qquad\qquad\quad~~\Bigl(\frac 1{2\beta}-\frac{i\kappa}{y}\Bigr)\,{\Phi}_2(y)=0\cr\spazio{1.4}\nonumber
&{}&\frac d{dy}\,{\Phi}_2(y)+\Bigl(\frac 1{2\beta}+\frac{i\kappa}{y}\Bigr)\,{\Phi}_1(y)+\cr   
&{}&\qquad\qquad\qquad\qquad i\Bigl(\frac 1{2\beta}+\beta({{E}}-y^2)\Bigr)\,{\Phi}_2(y)=0
\label{DiracEquationOmega}
\end{eqnarray}
By eliminating ${\Phi}_2(y)$  we find the second order equation
\begin{eqnarray}
&{}&\!\!\!\!\!\!\!{\frac {d^{2}}{dy^{2}}}\,{{\Phi} _{1}}(y)- {\displaystyle 
\frac {2\,i\,\kappa\,\beta  \,
}{y\,(y - 2\,i\,\kappa \,\beta )}}\,{\frac {d}{dy}}\,{{\Phi} _{1}}(y) \, + \spazio{1.2} \cr
&{}&\phantom{XX}
\Bigl( {{E}} - y^{2}- 
{\displaystyle \frac {\kappa ^{2}}{y^{2}}}+ 2\,i\,\beta \,y   + \beta^2\,({{E}}-y^{2} )^{2} - \spazio{1.2}\cr
&{}&\phantom{XXi}\frac\kappa y\,
{\displaystyle \frac {1 + 2\,\beta ^{2}\,({{E}} - y^{2} )}{(y - 2\,i\,\kappa \,\beta )}} \,  \Bigr)\,{{\Phi} _{1}}(y)=0 
\label{SecondOrderEquationOmega}
\end{eqnarray}

\smallskip\noindent
Besides infinity, this equation, with $y\in[\,0,+\infty)$, presents the obvious singularity at the origin,
absent in the one-dimensional case with $\kappa\!=\!0$ and
$y\in(-\infty,+\infty)$.
The non-relativistic limit for $\beta\rightarrow 0$, 
\begin{eqnarray}
\!\!\!\!\!\!{\frac {d^{2}}{dy^{2}}}\,{{\Phi}_1 }(y) + \Bigl[ \,{{E}} - y^{2}-\frac{\ell(\ell+1)}{y^2} \Bigr]\,{\Phi}_1(y)=0
\label{SecondOrderEquationOmega_limit}
\end{eqnarray}
reproduces the usual Schr\"odinger equation, with orbital angular momentum $\ell$, in a quadratic potential.
The properties of the equation with $\beta\ne 0$, however, are very different
from those of the non-relativistic counterpart, since (\ref{SecondOrderEquationOmega}), even neglecting the presence of an imaginary term,
 presents an asymptotic oscillatory behavior that prevents the existence of any normalizable solution. Therefore, as we said, the perturbation
expansion from the non-relativistic system turns out to be singular. 
 
We can transform the domain of definition of the differential equation  (\ref{SecondOrderEquationOmega}) with a map $U$ defined by
\begin{eqnarray}
 \!\!\!{\Phi}(y)=U({\Phi}_1(y))=\Bigl(1-\frac{2i\kappa\beta}y\Bigr)^{-1/2}\,{\Phi}_1(y).
\end{eqnarray}
The transformed differential equation becomes then 
\begin{eqnarray}
&{}&\!\!\!\!{\frac {d^{2}}{dy^{2}}}\,{{\Phi} }(y) + \Bigl[ \,{{E}} - y^{2}-\frac{\kappa(\kappa+1)}{y^2} +\spazio{1.0}\cr
&{}& 2\,i\,\beta \,\Bigl(y - {\displaystyle \frac {\kappa \,( \kappa+1 )}{(y - 2\,i\,\kappa \,\beta )\,y^{2}}} \Bigr)
 + \beta ^{2}\,\Bigl((y^{2} - {E} )^{2} +\spazio{1.4}\cr 
&{}&{\displaystyle \frac {2\,\kappa \,(y^{2} - {E} )}{(y - 2
\,i\,\kappa \,\beta )\,y}}  + {\displaystyle \frac {3\,\kappa ^{2
}}{(y - 2\,i\,\kappa \,\beta )^{2}\,y^{2}}} \Bigr)\, \Bigr]\,{\Phi}(y)=0\cr
&{}&
\label{SecondOrderEquationOmega_senza}
\end{eqnarray}

\smallskip\noindent
Observing that $\beta\!>\!0$, we find it useful to introduce the parameter $g\!=\!\mp i\beta$ according to whether $\kappa\!>\!0$ or $\kappa\!<\!0$ 
respectively. With such a choice of the signs, when $g\!>\!0$ no additional singularities besides the origin and the infinity appear at finite values of $y$ and the transformation $U$ is unitary. The equation to be studied can finally be written as 
\begin{eqnarray}
\!\!\! H_g\,{\Phi}(y)=0\,,\quad H_g \equiv-\frac{d^2}{dy^2}+V({{E}},g,\kappa,y)
\label{SecondOrderEquationOmega_g}
\end{eqnarray}
whose `potential' is given by
\begin{eqnarray}
V({{E}},g,\kappa,y)=V_0({{E}},\kappa,y)+V_1({{E}},\kappa,g,y) 
\label{Potential}
\end{eqnarray}
where
\begin{eqnarray}
&{}&\!\!\! V_0=y^2-{E}+\frac{|\kappa|(|\kappa|\pm1)}{y^2},\spazio{1.2}\cr
&{}& \!\!\!V_1=\pm\, 2\,g\,\Bigl(y - {\displaystyle \frac {|\kappa| \,(|\kappa| \pm 1)}{(y + 2\,
|\kappa| \,g)\,y^{2}}} \Bigr) +\spazio{1.2}\cr
 &{}& g^{2}\,\Bigl((y^{2} - {E} )^{2} \pm 
{\displaystyle \frac {2\,|\kappa| \,(y^{2} - {E} )}{(y + 2
\,|\kappa| \,g)\,y}}  + {\displaystyle \frac {3\,|\kappa| ^{2}}{(y + 
2\,|\kappa| \,g)^{2}\,y^{2}}} \Bigr)\spazio{1.2}\cr
\nonumber
\end{eqnarray}
The double signs correspond again to $\kappa\!>\!0$ and $\kappa\!<\!0$ respectively, while the dimensionless energy ${{E}}$,  entering $V({{E}},g,\kappa,y)$ in a polynomial expression, can be considered, for the moment, a parameter that will be determined during the calculation. It is common sense -- and it can also be proved rigorously -- that the equation  (\ref{SecondOrderEquationOmega_g}) with positive  $g$ admits bound states: starting from these we can then try to make an analytic continuation  in the complex plane of the coupling constant $g$ 
from the real positive to the positive or negative imaginary axis in order to recover the values of the parameters of the initial problem. The procedure, however, presents some delicate points. Indeed the analytic continuation of the
eigenvalue of the complete equation would be immediate if the eigenvalue itself was given as a convergent series expansion in the coupling constant $g$. Unfortunately this is not the case:  the perturbation expansion in $g$ is asymptotic and has zero radius of convergence, \cite{Dy} .
Since finally our original problem has exactly $g^2\!\!<\!0$, the system we are dealing with is therefore
close to an unstable quartic oscillator, but for some non polynomial terms for which an additional discussion is in order:
all the considerations we developed in the Introduction about the incompleteness, self-adjoint extensions and  
boundary conditions at infinity thus apply.

Let us now recall that the study of the asymptotic expansions for treating perturbation problems in quantum mechanics was systematically undertaken
since the beginning of the seventies and the main concept allowing to deal with divergent series was the Borel summability \cite{LASW,GG70,RS}.
Given a formal series 
\begin{eqnarray}
S=\sum_{n=0}^\infty \,a_n\, z^n
\label{seriegenerica}
\end{eqnarray} 
we define its Borel transform as 
\begin{eqnarray}
B(u)=\sum_{n=0}^\infty \,\bigl(\,a_n\,/\,\Gamma(n+1+\nu)\,\bigr)\,u^n
\label{BorelTransform} 
\end{eqnarray}
where $\nu$ is a fixed parameter independent of $n$ that is usually chosen in relation to the asymptotic behavior of the
coefficients $\{a_n\}$, but whose choice is otherwise irrelevant.
If then $B(u)$
has a nonvanishing radius of convergence, admits an analytic continuation to a neighborhood of the positive real axis and if the
integral 
\begin{eqnarray}
\!\!\!\!\!\!\!\!\!\! f(z)=z^{-1}\int_0^\infty du\,(u/z)^{\nu}\,\exp(-u/z)\,B(u)
\label{Borelint}
\end{eqnarray}
 converges in a so called ``Nevanlinna disk'' $C_R=\{{\mathrm{Re}}\, (z^{-1})>R^{-1},\,R>0\}$, we say that  $f(z)$ is
 the sum of $S$ in $C_R$. In some favorable cases, where stronger properties hold, more convenient summation methods have also been proposed 
as, for example,  the Stieltjes sum, \cite{SG73}: this can be defined when the classical Stieltjes moment problem has a solution and the 
result can be conveniently expressed in terms of a convergent Pad\'e approximation. There are many cases, however, in which the Borel
summability and, \textit{a fortiori}, the Stieltjes summability cannot be applied: notably, this occurs when the Borel transform develops some singular
point along the positive real axis, so that the integral (\ref{Borelint}) is not defined. 
 The DBS originates from the need of finding a more flexible method of summation that could be able to avoid the
difficulties connected with the emergence of the singularities. The idea is to follow closely the 
way used in the discussion of the Riemann-Hilbert boundary value problem with data assigned on the positive real axis. 
More precisely, we reduce the requirement of the analytic continuation of $B(u)$ in a neighborhood of the positive real axis 
to the assumption of the existence of the analytic continuation of $B(u)$
in the intersection of a neighborhood of the positive real axis with the upper complex half-plane, so that the boundary value $B(u+i0)$ is well defined for any $u>0$.
The integral yielding $f(z)$ is then replaced by 
\begin{eqnarray}
\phi(z)=z^{-1}\int_0^\infty  B(u+i0)\,\rho(u/z)\,du,
\label{uppersum}
\end{eqnarray}
where the measure $\rho(u)du$ is finite and has positive moments $\mu_k\!=\!\int_0^\infty u^k\rho(u)\, du$. 
The function $\phi(z)$ defined by the integral (\ref{uppersum}) is analytic in the upper half-plane and it is
called the \textit{upper sum}. 
In particular for $\mu_k\!=\!\Gamma(k+1+\nu)$ we have $\rho(u)\!=\!u^{\nu}\exp(-u)$ and  (\ref{uppersum}) reproduces the relationship
between boundary values of analytic functions and many distributions defined by the inverse Laplace transform, \cite{BW}.
Since then $B(u-i0)\!=\!\overline{B(u+i0)}$ we also define the \textit{lower sum}  
\begin{eqnarray}
\overline{\phi(\overline{z})}=z^{-1}\int_0^\infty \overline{B(u+i0)}\,\rho(u/z)\,du\,.
\nonumber
\end{eqnarray}
and we consider the real and the imaginary parts of $\phi(z)$:
\begin{eqnarray}
&{}& \!\!\!\!\!\!\!\!\!\!\!\!\!\!\!f(z)=\frac12\,\Bigl(\phi(z)+\overline{\phi(\overline{z})}\,\Bigr)={\mathrm{Re}}\,(\phi(z)),\spazio{1.0}\cr
&{}& \!\!\!\!\!\!\!\!\!\!\!\!\!\!\!d(z)=\frac12\,\Bigl(\phi(z)-\overline{\phi(\overline{z})}\,\Bigr)={\mathrm{Im}}\,(\phi(z)).
\label{fz_dz}
\end{eqnarray} 
It is then natural to assume $f(z)$ as the DBS of the series, while the discontinuity along the positive real axis, $d(z)$, 
is uniquely determined and it has a zero asymptotic power series expansion. In \cite{Ca} $d(z)$ was related to the Schwinger effective action.

\bigskip


\sect{Application of the DBS.}
\label{dbs}
\smallskip

In this section we study the large order perturbation approach to the equation (\ref{SecondOrderEquationOmega_g}) in  one and in three
space dimensions and we stress some physical relations between angular momentum and perturbation expansion. 
 For the sake of completeness we will recall some useful facts related to general and well known properties of operators: greater details can be found in \cite{K,RS,BS,GG78}. These notions can be collected in two different  main subjects. The first concerns the analytic properties of the operator family, related to the coupling constant of the perturbation part. This is essential if we want to explore the changes of the spectrum of the operators during the analytic continuation process. The second one deals with the estimates of the asymptotic growth of the perturbation series, that imply the existence of appropriate summation mechanisms.

The differential equation  defined in (\ref{SecondOrderEquationOmega_g}) and dependent upon a parameter $g$,  has nice 
properties for $g=0$, in the sense that we are able to determine exactly its discrete spectrum. The natural question to be posed is whether
an eigenvalue $\mu$ of $H_0\!=\!H_{g = 0}$ with a certain (algebraic) multiplicity $M$ gives rise to nearby eigenvalues $\mu_1(g),...,\mu_s(g)$ with total multiplicity $M$ when the perturbation is switched on. If this is the case,  the eigenvalue $\mu$ is said to be stable. We just recall
that the discussion of this type of problems is  simple enough when the perturbation  potential $V_1({{E}},\kappa,g,y)$ is $H_0$-bounded, namely when the domain of definition ${\mathcal D}(V_1)$  contains ${\mathcal D}(H_0)$, since this property yields  $\|V_1\phi\|\leq a\|H_0\phi\|+b\|\phi\|$, for constant $a,\,b$ and for any $\phi\in{\mathcal D}(H_0)$. From the previous inequality, as in the proof of the Kato-Rellich theorem, 
we can easily deduce that ${\mathcal D}(H_g)$ is independent of $g$ and that, for any $\phi\in{\mathcal D}(H_g)$, $H_g\phi$ is an entire analytic function of $g$. In this case the perturbation scheme is said to be regular.
A slightly generalized strategy must be adopted when the perturbation is not $H_0$-bounded and specially when the behavior of the whole system depends dramatically upon the sign of the coupling constant or its square, as in our specific problem.
The idea is to extract the fundamental properties of the
relatively bounded case by  means of the notion of analytic family of operators, introduced by Kato \cite{K}: this term denotes a collection of operators $\{H_g\}_{g\in R}$ dependent upon a coupling constant $g$ taking values in a region $R$ of the complex plane, where each $H_g$ has a non-empty resolvent and is defined on a domain ${\mathcal D}$ independent of $g$; moreover, for each $\phi\in{\mathcal D}$, $H_g\phi$ is required to be strongly analytic in $g$. Although weaker than relative boundedness, this analyticity property is still sufficient to study 
the continuation of the eigenvalues in a region of the complex plane of $g$ and it can be proved that for $\xi$ in an appropriate open region of the resolvent set of $H_{g_{{}_0}}$ the resolvent $(\xi-H_{g})^{-1}$ is analytic in
$g$  when $|g-{g_{{}_0}}|$ is sufficiently small. The stability of discrete and non-degenerate eigenvalues immediately follows. For isolated eigenvalues of multiplicity $n$ there exist $\ell$ families of eigenvalues $\lambda_i(g)$  admitting  Puiseux expansions in terms of $g^{1/q_i}$, for  integers $\{q_i\}_{i=1,\ell}$, with total multiplicity $n$. We are therefore left with the need to prove that a family of operators is analytic:  the technique commonly used for this purpose is to establish appropriate quadratic estimates that allow us to deduce that each operator of the family is closed and that the perturbation term is small -- in the sense of Kato  \cite{K} -- with respect to any member of the family,  so that the domains are indeed independent of the coupling constant. This argument can be formalized as follows \cite{RS}: if $T_0$ and $T_1$ are closed operators with ${\mathcal D}(T_0)\cap{\mathcal D}(T_1)$ dense, then $T_0+T_1$ is closed if the inequality
\begin{eqnarray}
&{}&\!\!\!\!\!\!\!\!\!\! \Bigl(T_0+T_1\Bigr)^*\,\Bigl(T_0+T_1\Bigr)\geq a\Bigl(T_0^*\,T_0+T_1^*\,T_1\Bigr)+b\spazio{1.0}\cr
&{}&
\label{closure_cnd}
\end{eqnarray}
is satisfied for some constants $a,\,b$. The  one-dimensional quartic oscillator 
with a Hamiltonian $H\!=\!p^2\!+\!x^2\!+\!g^2x^4$
describes a situation very close to the Dirac equation in one space dimension we want to discuss and its specific estimate, \cite{BS}, can be formulated  as follows:
for  $a\!<\!1\!-\!|{\mathrm{Re}}(g^2)|/|g^2|$ and for all $\phi\in{\mathcal D}(p^2)\cap{\mathcal D}(x^4)$ there exists $b$ such that
$a\,(\|(p^2\!+\!x^2)\phi\|^2\!+\!|g|^4\,\|x^4\phi\|^2)\leq\|(p^2\!+\!x^2\!+\!g^2 x^4)\phi\|^2\!+\!b\,\|\phi\|^2$.
We finally observe that this estimate implies
the Herglotz property of $\lambda_i(g^2)$, namely ${\mathrm{Im}} (\lambda_i(g^2))>0$ for ${\mathrm{Im}} (g^2)>0$, that
gives precise informations about the analytic structure of $\lambda_i(g^2)$
and shows that $g=0$ cannot be an isolated singular point, but it must be a limit point of singularities of  the eigenvalues. 

We will now discuss the application of these general facts to the Dirac equation in a quadratic vector potential.
\bigskip

\noindent $(a)$ \textit{The case in one space-dimension.}
\bigskip 

Let us consider first  the one-dimensional problem. If we put $\kappa\!=\!0$ in (\ref{SecondOrderEquationOmega_g}$\,$-\ref{Potential}), we have the differential equation 
\begin{eqnarray}
 &{}&\!\!\!\!\!\!\!\!\! \Bigl[\,p^2+ (1-2{E}g^2)y^{2} +2gy+g^2y^4
 \,\Bigr]{\Phi}(y)\phantom{XXX}\spazio{1.2}\cr
 &{}& \phantom{XXXXXXXx}= {E}(1-g^2{E})\,{\Phi}(y)
\label{equa1dim_g}
\end{eqnarray}
defined for $-\infty\!<\!y\!<\!\infty$, where $p\!=\!-i\,(d/dy)$. As we said, in order to apply the general
theory, we need a quadratic estimate analogous to that of the anharmonic oscillator.
A less cumbersome notation is obtained if we rescale the variables by a dilation $y\rightarrow \alpha y$,
$p\rightarrow p/\alpha$ and then we choose 
\begin{eqnarray}
&{} \lambda={E}\alpha^2(1-{E}g^2) \,,\qquad\sigma=2g\alpha^3\,,\spazio{1.0}\cr 
&{} \alpha=(1-2{E}g^2)^{-1/4}\,.
\label{alpha_sigma}
\end{eqnarray} 
The inverse relations are
\begin{eqnarray}
&{} {E}=2\lambda\,(1+\lambda\sigma^2)^{1/4}\,\bigl[\,1+(1+\lambda\sigma^2)^{1/2}\,\bigr]^{-1}\spazio{1.0}\cr
&{} g={\displaystyle \frac\sigma 2}\,(1+\lambda\sigma^2)^{1/8}\,,
\label{alpha_sigma_inverse}
\end{eqnarray} 
and by means of (\ref{alpha_sigma_inverse}) it is possible to reconstruct the function ${E}(g)$ from $\lambda(\sigma)$.
In the new parametrization (\ref{alpha_sigma}), equation (\ref{equa1dim_g}) becomes
\spazio{-1.}
\begin{eqnarray}
\Bigl[\,p^2+y^2+\sigma y+\frac{\sigma^2}{4}y^4\,\Bigr]{\Phi}(y)=\lambda\,{\Phi}(y)
\label{EquaD-1dim_lambda}
\end{eqnarray}
and the quadratic estimate has to be done for the operator on the left hand side of (\ref{EquaD-1dim_lambda}).
For $\phi\in {\mathcal D}(p^2)\cap{\mathcal D}(y^4)$ and ${\mathrm{Im}} (\sigma)\not=0$ we consider
the closed operators
\begin{eqnarray}
&{}&T_0=p^2+y^2+{\mathrm{Re}}(\sigma) y\,,\spazio{0.8}\cr
&{}&T_1=i\,{\mathrm{Im}}(\sigma)y+\frac{\sigma^2}{4}y^4\,.
\label{H0H1_1dim}
\end{eqnarray}
Everything is known about  $T_0$ since it is just the Hamiltonian of a harmonic oscillator in the displaced variable
$(y+{\mathrm{Re}}(\sigma)/2)$.
A direct calculation  given in Appendix A proves the quadratic estimate (\ref{closure_cnd}).

It will be shown in Section \ref{numeri} that for parity reasons only even powers of the coupling constant
$\sigma$ contribute to the perturbation expansion of the eigenvalues.
As a consequence of (\ref{quadratic_estimate_1dim})  we can state that on the domain 
${\mathcal D}(p^2)\cap{\mathcal D}(x^4)$ and with $T_0,T_1$ given in (\ref{H0H1_1dim}),
the operators $H_\sigma=T_0+T_1$ form a holomorphic family  with compact resolvents for $\sigma^2$ in the complex plane cut along the negative axis. We
thus  get a perturbations series for each eigenvalue
\begin{eqnarray}
\lambda(z)=\sum\limits_{n=0}^{\infty}~a_n\,z^{n}
\label{seriepertgenerale}
\end{eqnarray}
with $z=\sigma^2$. For (\ref{seriepertgenerale}) there exists a Borel transform (\ref{BorelTransform}). 
The corresponding sum is given by
\begin{eqnarray}
\!\!\!\!\!\!\!\!\!\!\lambda(z) = z^{-1-\nu}\,\int\limits_{0}^{\infty}\,u^{\nu}\,\exp(-u/z)\,B(u+i0)\,du
\label{BorelInversa}
\end{eqnarray}

\bigskip

\noindent $(b)$ \textit{The case in three space-dimensions.}
\medskip 

We next consider the three-dimensional problem assuming $\kappa\!=\!-1$ in order to simplify the notation:  indeed the treatment is qualitatively the same for any other allowed value of $\kappa$. The equation (\ref{SecondOrderEquationOmega_g}, \ref{Potential}), then,  specifies to
\begin{eqnarray}
&{}&\!\!\!\!\!\!\!\! H_g\,{\Phi}(y)\equiv
\Bigl[\,p^2 \! +\! y^{2}\! -\! 2\,g\,y \!+\! g^{2}\,\Bigl((y^{2}\! - \!{E})^{2}-{\displaystyle \frac {2\,(y^{2}\! -\! {E})}{y\,(y\! +\! 2\,g)}}\!\spazio{1.0}\cr
&{}&\phantom{XXXX}
  + 
{\displaystyle \frac {3}{y^{2}\,(y \!+\! 2\,g)^{2}}} \Bigr) \,\Bigr]{\Phi}(y)={E}\,{\Phi}(y)
\label{equa3dim_g_k0} 
\end{eqnarray}
and since $g$ is proportional to $c^{-1}$ the limit $g\rightarrow 0$ correctly reproduces the Schr\"odinger equation (\ref{SecondOrderEquationOmega_limit}) with angular momentum $\ell=0$.

With respect to (\ref{equa1dim_g}), equation (\ref{equa3dim_g_k0}) contains non polynomial terms proportional to $g^2$.
The rational terms have vanishing denominators when both $g$ and $y$ tend to zero and the simultaneous limit is not well defined. 
The expansion in $g$ of the perturbation potential leads to terms with higher and higher divergences in the origin:
\begin{eqnarray}
&{}&\!\!\!\!\!V_1\simeq -2gy+g^{2}\Bigl[{y^{-4}}\bigl(  3 + 2\,{E}\,y^{2} + ({E}^{2} - 2)\,y^{4}\cr
&{}&\!\!\!\!\! - 2\,{E}\,y^{6} + y^{8}\bigr)  - g{y^{-5}}\bigl(\,{\displaystyle  {4\,(3 + 
{E}\,y^{2} - y^{4})}}\bigr) 
 + \mathrm{O}(g^{2})\Bigr] 
\nonumber
\end{eqnarray} 
These terms make rather problematic the possibility of achieving a reasonable quadratic estimate.
On the other hand the expansion in $y$ in the neighborhood of the origin gives, up to to terms of $O(y)$,
\begin{eqnarray}
&{}&\!\!\!\!\!\!\!\!\!\!V_1\simeq {\displaystyle \frac {3}{4y^2}}  - {\displaystyle \frac { 3
 - 4\,{E}\,g^{2}}{4\,gy}}+
{\displaystyle \frac {9 - 8\,{E}\,g^{2
} + 16\,{E}^{2}\,g^{4}}{16\,g^{2}}}\cr 
 &{}&
\label{V1}  
\end{eqnarray} 
As the quadratic divergence
at the origin is due to the angular momentum of the system, the term $(3/4)\,y^{-2}$  
shows that, as a consequence of relativity, we are dealing with a spin $(1/2)$ fermion and that the spin  is the only angular momentum left when
 $\kappa\!=\!-1$. Besides these physical observations, however, the expansion cannot
be used perturbatively, not even for searching the local solutions at the origin. A different approach has therefore to be searched.
We thus find it necessary to fix the final value of the parameter $g$ and to put this final value $g_*$ in the singular
 terms: we say that we \textit{freeze} the parameter at its final value wherever it is needed. 
We then write the equation 
\begin{eqnarray} 
&{}&\!\!\!\!\!\!\!\!\Bigl(\Bigl[\,p^{2} + y^{2}  + {\displaystyle \frac {3}{4\,y^{2}}}\Bigr]  -\Bigl[\, 2\,g\,y  
+g^2\,\Bigl(\,\frac {3}{4\,g^2_\ast}\,\frac{y+4g}{(y+2g_\ast)^2\,y}
  \spazio{1.2}\cr
&{}&
- (y^{2} - {E})^{2} + 2\,{\displaystyle \frac {y^{2}-{E}
 }{(y + 2\,g_\ast)\,y}}\Bigr)\Bigr]\,\Bigr)\,{\Phi}(y)={E}\,{\Phi}(y)\cr
&{}&
\label{NewEq3dim}
\end{eqnarray} 
that coincides with  (\ref{equa3dim_g_k0}) when $g_\ast\!=\!g$.
We denote respectively by $T_0$ and $T_1$ the operators in the first and in the second square bracket; we then
observe that with $g$ frozen at any non-vanishing $g_\ast$ the factors $(y+2g_\ast)^{-1}$ are bounded in $0\!\leq \!y\!\leq \!+\infty$ and do not need further
 estimates.
Therefore the  two terms of $T_1$ diverging like $y^{-1}$ at the origin and presenting an at most constant behavior at infinity are small in the sense of Kato with respect to $y^2+(3/4)\,y^{-2}$ in $T_0$, so that they can be neglected in the estimate. Since it
is well known that $p^2+y^2+j(j+1)y^{-2}$ is a closed operator on the maximal domain of the functions satisfying
the condition ${\Phi}(y)\simeq y^{j+1}$ at $y=0$, that correspond to the regular solutions of the differential equation (\ref{equa3dim_g_k0})  --  and therefore $T_0$ is closed on the maximal domain with
${\Phi}(y)\simeq y^{3/2}$ at the origin  -- , by means of a calculation analogous to 
(\ref{quadratic_estimate_1dim}) we conclude that $T_0+T_1$ is closed and that the set $\{H_g\}$, whose elements are specified in (\ref{equa3dim_g_k0}), form an analytic family of operators with compact resolvent. We can thus conclude that 
for the three-dimensional case also we get a perturbation series like (\ref{seriepertgenerale}) with $z=\sigma^2$ for which
the Borel transform (\ref{BorelTransform}) and its inverse  (\ref{BorelInversa}) are well defined.

It is evident that the concrete calculation of the perturbation expansion 
is much more difficult in three dimensions than in one. In practice 
the expansion (\ref{V1}) could suggest as preferable the use of a variational method by defining the isospectral dilated operator
\begin{eqnarray} 
&{}&\!\!\!\!\!\!\!\!\!\!\!\!\!H_{\eta}(g,{E})={\eta}^{-2}\Bigl(\,\Bigl[\,p^{2} + y^{2}  + {\displaystyle \frac {3}{4\,y^{2}}}  
\,\Bigr]\spazio{1.2}\cr   
&{}&+\Bigl[\,({\eta}^4-1)y^{2}
-{\displaystyle{\eta}\frac {3\,(3{\eta} y + 4\,g)}{4\,({\eta} y + 2\,g)^2\, y}} 
- 2\,g\,{\eta}^3 y\spazio{1.2}\cr
&{}&+ g^{2}{\eta}^2\,({\eta}^2y^{2} - {E})^{2}
+ 2g^2{\eta}\,{\displaystyle \frac {{E}
 - {\eta}^2y^{2}}{({\eta} y + 2\,g)\, y}} \,\Bigr]\,\Bigr)
\nonumber
\label{Dilated3dim}
\end{eqnarray} 
where $g=-i|g|$, ${\eta}=\exp(i\vartheta)$, $0<\vartheta<\pi/6.$ 
We could then approximate the first eigenvalues restricting the Hamiltonian on the linear space of the first $n$ eigenvectors of
$\,p^{2} + y^{2}  + (3/4)\,y^{-2}$. We leave this problem for future investigations.

\bigskip


\sect{Numerical developments.}
\label{numeri}
\smallskip

In this last section we present the numerical calculations concerning the perturbation treatment of equation (\ref{EquaD-1dim_lambda})
yielding the determination of $\lambda(\sigma)$ from which we can obtain ${E}(g)$ by inverting the definitions (\ref{alpha_sigma}).
We take $p^2+y^2$ as unperturbed operator and we write the perturbation part in the form $\sigma U_1+\sigma^2U_2$, where
$U_1=y$ and $U_2=y^4/4$. We introduce the usual creation and destruction operators
\begin{eqnarray}
&{}a=2^{-1/2}\,\bigl({d}/{dy}+y\bigr)\,,
\quad  \ad=2^{-1/2}\,\bigl(-{d}/{dy}+y\bigr)\,,\spazio{1.0}\cr
&{}\bigl[\,a,\ad\,\bigr]=1\,,
\label{a-ad}
\nonumber
\end{eqnarray} 
so that
\begin{eqnarray}
&{} p^2+y^2=2\ad a+1,\qquad y=2^{-1/2}\,(\ad+a),\spazio{1.0}\cr
&{} y^4=4^{-1}\,({\ad}^4+4{\ad}^3a+6{\ad}^2a^2+4\ad a^3\spazio{0.7}\cr
&{}+a^4+6{\ad}^2+12\ad a+6a^2+3).
\label{y2y4}
 \nonumber
\end{eqnarray} 
In the standard Dirac notation, the solutions of the eigenvalue equation for the unperturbed operator, $(2\ad a\!+\!1)\ket\psi=\lambda\ket\psi$,
are the usual occupation number states $\ket n$, expressed in terms of Hermite functions, where $\lambda\!=\!\lambda_n\!\equiv \!2n\!+\!1$ with integer $n$.
For later use we recall the matrix elements of $y$, $y^2$ and $y^4$ with respect to pairs of such states, namely
\begin{eqnarray}
&{}&\!\!\!\!\!\matrel k{\,y\,}n=2^{-1/2}\Bigl(\sqrt{n\!+\!1}\,\delta_{k,n+1}+\sqrt{n}\,\delta_{k,n-1}\Bigr)\spazio{1.4}\cr
&{}&\!\!\!\!\!\matrel k{y^4}n=2^{-2}\Bigl(\sqrt{(n\!+\!4)(n\!+\!3)(n\!+\!2)(n\!+\!1)}\,\delta_{k,n+4}+\spazio{0.8}\cr
&{}&(4n\!+\!6)\sqrt{(n\!+\!2)(n\!+\!1)}\,\delta_{k,n+2}\,+\spazio{0.8}\cr
&{}&(6n^2\!+\!3(2n\!+\!1))\,\delta_{k,n}+
(4n\!-\!2)\sqrt{n(n\!-\!1)}\,\delta_{k,n-2}\,+\spazio{0.8}\cr
&{}&\sqrt{n(n\!-\!1)(n\!-\!2)(n\!-\!3)}\,\delta_{k,n-4}\Bigr)
\label{matrel}
\end{eqnarray} 
where $\delta_{i,j}$ is the usual Kronecker symbol, taking values one or zero according to whether the indices are equal or different. Since the Hermite functions have the parity of $n$, we see from (\ref{matrel}) that
$\matrel k{U_1}n$ is vanishing when the integers $k$ and $n$ have the same parity and $\matrel k{U_2}n$ is vanishing when the parity
is opposite. We now set up the standard perturbation framework by defining 
\begin{eqnarray}
&{}\lambda_n=(2n+1)+\sum\limits_{i=1}^{\infty}g^i \lambda_n^{(i)}\cr
&{}\ket{\psi_n}=\ket n +\sum\limits_{i=1}^{\infty}g^i \pert i,\,~ .
\nonumber
\end{eqnarray} 
As usual, the global phase of $\ket{\psi_n}$ can be chosen in such a way to have $\scal n{\psi_n^{(1)}}=0$. A straightforward calculation leads then to the first and second order quantities
\begin{eqnarray}
 &{}& \lambda_n^{(1)}=0\,,\quad \lambda_n^{(2)}=2^{-4}\bigl(6n(n\!+\!1)\!-\!1\bigr)\,,\spazio{0.6}\cr
&{}& \pert 1=-(2\sqrt2)^{-1}\bigl(\sqrt{n\!+\!1}\,\,\ket{n\!+\!1}-\sqrt{n}\,\,\ket{n\!-\!1}\bigr)\,,\spazio{1.0}\cr 
&{}&\pert 2=-\frac{\sqrt{(n\!+\!4)(n\!+\!3)(n\!+\!2)(n\!+\!1)}}{128}\,\ket{n+4}\cr
&{}&\phantom{\pert 2}+\frac{(2n\!+\!1)\sqrt{(n\!+\!2)(n\!-\!1)}}{32}\,\ket{n+2}\cr
&{}&\phantom{\pert 2}+\frac{(2n\!+\!1)\sqrt{n(n\!-\!1)}}{32}\,\ket{n-2}\cr
&{}&\phantom{\pert 2}+\frac{\sqrt{n(n\!-\!1)(n\!-\!2)(n\!-\!3)}}{128}\,\ket{n-4}
\label{firstorder}
\end{eqnarray} 
and to the recurrence relation, for $i>2$,
\begin{eqnarray}
&{}&\!\!\!\!\!\!\!\!\!\!\!\!\!\!\! \Bigl[(p^2+y^2)-(2n+1)\Bigr]\,\pert i+U_1\,\pert{i-1}+\cr
&{}&U_2\,\pert{i-2} -\sum\limits_{s=1}^{i-1}\lambda_n^{(s)}\,\pert{i-s}-\lambda_n^{(i)}\ket n=0\,.\cr
&{}&
\label{recurrence}
\end{eqnarray} 

We now prove by induction that for any $m\geq 0$ we have: 
\bigskip

$(a)$ $~\lambda_n ^{(2m+1)}\!=0$, 
\smallskip

$(b)$ $~\pert{2m+1}$ has the opposite parity of $n$, 
\smallskip

$(c)$   $~\pert{2m+2}$ has the same parity of $n$. 

\bigskip\noindent
The relations (\ref{firstorder})  give the initial step of the inductive argument. Suppose now the properties we require are true for $i\leq 2m$. The scalar product of (\ref{recurrence}) by $\ket n$ gives 
\begin{eqnarray}
 \lambda_n^{(i)}=\matrel n{U_1}{\psi_n^{(i-1)}}+\matrel n{U_2}{\psi_n^{(i-2)}}
\label{lambda_i}
\end{eqnarray} 
and we first consider $i=2m+1$. By the inductive hypothesis both the matrix elements of $U_1$ and $U_2$ are vanishing. The last sum is also vanishing,
since $ \lambda_n^{(s)}\!=0$ for odd $s$ and $\scal n{\psi_n^{(2m+1-s)}}=0$  for even $s$. As a consequence $\lambda_n^{(2m+1)}\!=0$. Looking then at the perturbation contributions to the states, we have
\begin{eqnarray}
 \ket{\psi_n^{(i)}}=\sum\limits_{k=0}^{\infty}\ket k\scal k{\psi_n^{(i)}}
\label{psi_i}
\end{eqnarray} 
where the scalar product of (\ref{recurrence}) by $\ket k$ provides the relation, for $k\not=n$,
\begin{eqnarray}
&{}\!\!\!\!\!\scal k{\psi_n^{(i)}}=(2n\!-\!2k)^{-1}\Bigl( \matrel k{U_1}{\psi_n^{(i-1)}}+\phantom{XX}\cr
&{}\matrel k{U_2}{\psi_n^{(i-2)}}-\sum\limits_{s=1}^{i-1} \lambda_n^{(s)}\scal k{\psi_n^{(i-s)}}\Bigr)
\label{coeff_psi_i}
\end{eqnarray} 
from which the parity properties are easily deduced. 
The induction is therefore complete and it implies, in particular, that the perturbation expansion of the eigenvalues is in $\sigma^2$ rather that in $\sigma$.

By using  (\ref{matrel}-\ref{coeff_psi_i}) we can set up a recursion scheme by which the expansions of the eigenvalues are determined. With the help of  some computer algebra we have calculated the first 94 coefficients of the perturbation expansion in 
$\sigma$ of the lowest unperturbed eigenvalue and of the fundamental state
in terms of rational numbers, namely with infinite precision. Starting from there we have continued the expansion with a fortran code in floating point up to 250th order,  thus obtaining a contribution to the eigenvalue up to order 125 in $\sigma^2$. 
The coefficients of the asymptotic series in $\sigma^2$ have alternate signs and,  on the basis of the numbers we have calculated, we find the evidence for an asymptotic behavior
\begin{eqnarray}
 \!\!\!\!\!\!\!\!\!\!\!\! a_n\approx \frac\pi 2\,\Bigl(1-\frac{\sqrt{6}}{\pi^{3/2}}\Bigr)\,(-1)^n\,\Bigl(\frac 38\Bigr)^n\,\Gamma(n+1/2)
\label{a_n_asymptotic}
\end{eqnarray}
To be more precise, we have defined the sequence $\{\breve{a}_n\}$ where $\breve{a}_n$ is given by the ratio of the coefficient $a_n$ of the initial divergent series divided by the right hand side of equation (\ref{a_n_asymptotic}). We have then used the Shanks transformation  $A_n=(\breve{a}_{n+1}\breve{a}_{n-1}-\breve{a}_n^2)/(\breve{a}_{n+1}+\breve{a}_{n-1}-2\,\breve{a}_n)$,  \cite{sidi}, in order to improve the convergence of $\{\breve{a}_n\}$ and we have found $A_{125}=1.00009$, where the rounding numerical error can be estimated of the order of $10^{-4}$.

Some comments on the general features of the perturbation series and of its Borel transform are in order. For positive values of $\sigma^2$ the series is Borel summable. 
For negative $\sigma^2$ (namely for
imaginary $\sigma$, which is the case we are interested in), all the terms of the series acquire equal signs and the series itself becomes Borel summable only in the distributional sense. 
We will therefore assume the asymptotic series with all positive terms and thus consider
$z=\sigma^2=4\,\Omega_*>0$ in the Borel anti-transformation integral (\ref{BorelInversa}): the meaning of $\Omega_*$ will be clarified below.
Dropping $(-1)^n$ in the asymptotic behavior (\ref{a_n_asymptotic}) it is clear that the Borel transform develops singularities on the positive real axis, the first of which has to be expected in the neighborhood of 8/3. The shift in the energy of the state, given by $f(z)$ in (\ref{fz_dz}), can be plainly calculated by taking the principal part of 
(\ref{BorelInversa}) with respect to the poles on the real axis. 

The imaginary part of the perturbed energy, given by $d(z)$ in (\ref{fz_dz}), has a much more
interesting physical meaning: indeed $2\,d(z)$ gives the decay constant $\Gamma$ of the state itself as in (\ref{normdecay}) and, in an interpretation connected with a second quantization framework, to the particle-antiparticle production rate 
$w^f=2\,{\mathrm{Im}}{\mathcal L}_{\mathrm{eff}}$, where ${\mathcal L}_{\mathrm{eff}}$
is the effective Lagrangian of the system, \cite{Sch,GS}. The imaginary part will be computed by summing the contributions of the positive real poles to the integral (\ref{BorelInversa}), calculated along the integration path encircling those poles in the upper complex half plane. 
An efficient way to make the calculation is to use a Pad\'e approximation for
the Borel transform $B(u)$ of the perturbation series. Since $B(u)$ is expressed by a series expansion up to order 125, we
can take Pad\'e approximants of rather high order and to look at the stabilization of the values of the poles. The results for the seven lowest poles are summarized here below.

\medskip
\begin{small}
\begin{center}
\begin{tabular}{c|c|c|c|c|c}
\texttt{[58,59]}  & \texttt{[59,58]}  & \texttt{[59,59]} & \texttt{[59,60]}  & \texttt{[60,59]}  & \texttt{[60,60]}\\ 
\hline
\hline
\texttt{2.67561} & 
\texttt{2.67562} &
 \texttt{2.67571} & \texttt{2.67575} & \texttt{2.67576} &
 \texttt{2.67572}\\
\texttt{2.71528} & \texttt{2.71529} &
 \texttt{2.71583} & \texttt{2.71611} & \texttt{2.71612} &
 \texttt{2.71589}\\
\texttt{2.79311} & \texttt{2.79314} &
 \texttt{2.79480} & \texttt{2.79563} & \texttt{2.79564} &
 \texttt{2.79497}\\
\texttt{2.92226} & \texttt{2.92235} &
 \texttt{2.92678} & \texttt{2.92899} & \texttt{2.92901} &
 \texttt{2.92722}\\
\texttt{3.12725} & \texttt{3.12752} &
 \texttt{3.13965} & \texttt{3.14582} & \texttt{3.14589} &
 \texttt{3.14082}\\
\texttt{3.45803} & \texttt{3.45883} &
 \texttt{3.49492} & \texttt{3.51474} & \texttt{3.51500} &
 \texttt{3.49830}\\
\texttt{4.03675} & \texttt{4.03958} &
 \texttt{4.17104} & \texttt{4.26976} & \texttt{4.27146} &
 \texttt{4.18431}\\
\end{tabular} 
\end{center}
\label{tavola1}
 \end{small}
\medskip

 From these numbers it appears, first of all, that the locations of the lowest poles 
are in the neighborhood of $8/3$, as deduced from (\ref{a_n_asymptotic}): this was to be expected and indicates the consistency of the approximation, as  
we said in the Introduction.
We then see that their values
tend to stabilize with the increasing order of the Pad\'e, the lower poles obviously faster than the upper ones, for which approximants of higher order would be required. Therefore, recalling that no further information can be deduced by the Pad\'e construction  when the sum of the degrees of numerator and denominator exceeds the order of the series, a considerably larger number of perturbation terms should be calculated in order to stabilize the higher poles. This means, in turn, that the calculations of the real and imaginary parts of the perturbed energy give much more precise results for small values of $z$, for which the contribution of the lowest poles is largely dominant: a natural fact in a perturbation framework, that can be clearly observed in the next table,

\medskip
\begin{small}
\begin{center}
\begin{tabular}{c|c|c|c|c}
${\Omega}$  
& \texttt{[57,57]}  & \texttt{[58,58]} & \texttt{[59,59]}  & \texttt{[60,60]}   \\ 
\hline\hline
 \texttt{0.05} & 
\texttt{.251588e-5} &
\texttt{.251584e-5} & \texttt{.251588e-5} & \texttt{.251586e-5} 
 \\
 \texttt{0.08} & 
\texttt{.321991e-3} &
\texttt{.321360e-3} & \texttt{.321933e-3} & \texttt{.322124e-3} 
 \\
 \texttt{0.10} & 
\texttt{.162741e-2} &
\texttt{.161602e-2} & \texttt{.162615e-2} & \texttt{.163092e-2} 
 \\
 \texttt{0.12} & 
\texttt{.483648e-1} &
\texttt{.476170e-1} & \texttt{.482741e-1} & \texttt{.486376e-1} 
 \\
 \texttt{0.15} & 
\texttt{.146167e-1} &
\texttt{.141402e-1} & \texttt{.145544e-1} & \texttt{.148164e-1} 
 \\
 \texttt{0.20} & 
\texttt{.456068e-1} &
\texttt{.426354e-1} & \texttt{.451921e-1} & \texttt{.470098e-1} 
 \\
\texttt{0.25} & 
\texttt{.925871e-1} &
\texttt{.837260e-1} & \texttt{.913049e-1} & \texttt{.970549e-1} 
 \\
\end{tabular} 
\end{center}
\label{tavola2}
\end{small}
\medskip

\noindent
where, for different values of ${\Omega}$,  we have reported the numerical data for the imaginary 
part of (\ref{BorelInversa}), namely
\begin{eqnarray}
&{}&{\mathrm{Im}}(\lambda)=\displaystyle{-\frac{\pi}{({4_{\,}{\Omega_*}})^{1+\nu}} }\,\sum\limits_{\{\mathrm{poles}\,\,p_i\}} {p_i}^{\nu}\,\phantom{XXXXXXX} \cr
&{}&{}\qquad{\mathrm{Res}}_{\,p_i}\Bigl(P_{\,[m,n]}(B(u))\Bigr)\,\,
\exp\Bigl({\displaystyle{-\frac{p_i}{4_{\,}{\Omega_*}}}}\Bigr)\,
\label{residui}
\end{eqnarray}
with $\nu=-1/2$.
In (\ref{residui})  $B(u)$ is the Borel transform (\ref{BorelTransform}) of the perturbation series and
$P_{\,[m,n]}$ is the $[m,n]$ Pad\'e approximant of $B(u)$ with poles $\{p_i\}$. 
By ${\mathrm{Res}}_{\,p_i}$ we have indicated, as usual, the residue at the pole $p_i$.
Finally the value of $\Omega_*$ has been constructed by using the formulas (\ref{alpha_sigma}) with 
$\Omega_*=\Omega\,(1+2{E}\Omega)^{-3/2}$ and $\Omega=|g|^2$ is the same parameter used in \cite{GS}.
We strongly stress that while in (\ref {alpha_sigma}), with imaginary $g$ and $\sigma$,  
$\lambda=E$ to the order $|g|^2$, the relationship between $\sigma$ and $g$ is not so straightforward, as the correction due to the term $\alpha^6$ is large
and cannot by no means be neglected: on the contrary, it proves fundamental to establish
the correct relationship between the perturbation data and the data calculated in \cite{GS},
taking the zeroth order $\lambda=1$ inside $\alpha$.
 
We also report a further check of stability of the results: the numerical values
give the imaginary part (\ref{residui}) calculated with the use of the $[60,60]$ Pad\'e approximant for different values of the constant $\nu$.

\medskip
\begin{small}
\begin{center}
\begin{tabular}{c|c|c|c|c}
${\Omega}$  
& \texttt{$\nu\,$=-1/2}  & \texttt{$\nu\,$=0} & \texttt{$\nu\,$=1/2}  & \texttt{$\nu\,$=1}   \\ 
\hline\hline
 \texttt{0.05} & 
\texttt{.251586e-5} &
\texttt{.251586e-5} & \texttt{.251586e-5} & \texttt{.251586e-5} 
 \\
 \texttt{0.08} & 
\texttt{.322124e-3} &
\texttt{.322124e-3} & \texttt{.322099e-3} & \texttt{.322113e-3} 
 \\
 \texttt{0.10} & 
\texttt{.163092e-2} &
\texttt{.163309e-2} & \texttt{.163333e-2} & \texttt{.163335e-2} 
 \\
 \texttt{0.12} & 
\texttt{.486376e-1} &
\texttt{.488771e-1} & \texttt{.489565e-1} & \texttt{.489141e-1} 
 \\
 \texttt{0.15} & 
\texttt{.148164e-1} &
\texttt{.150256e-1} & \texttt{.151061e-1} & \texttt{.150456e-1} 
 \\
 \texttt{0.20} & 
\texttt{.470098e-1} &
\texttt{.485809e-1} & \texttt{.487832e-1} & \texttt{.484381e-1} 
 \\
\texttt{0.25} & 
\texttt{.970549e-1} &
\texttt{.101983} & \texttt{.100359} & \texttt{.100484} 
 \\
\end{tabular} 
\end{center}
\label{tavola3}
\end{small}
\medskip

\begin{figure}
\begin{center}
\includegraphics*[height=4.8 cm,width=7.cm]{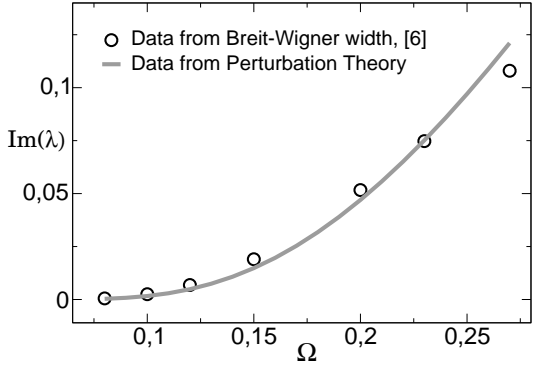}
\caption{Imaginary part of the perturbed lowest eigenvalue.
\textit{Dotted line}: perturbation data. \textit{Solid line}: data from \cite{GS}. }
\label{prod}
\end{center}
\end{figure}
We will conclude the section by comparing  the results here obtained with those presented in \cite{GS}. The methods used to get them, as we said, are completely different. On the one hand, we have a perturbation approach that becomes more and more precise when the perturbation parameter decreases and needs many further terms for intermediate values of that parameter: the imaginary part of the energy is directly calculated by using the Pad\'e approximants to invert the Borel transform in the distributional sense and integrating along the appropriate path in the complex plane. On the other, the determination of the spectral concentration is obtained by a numerical integration of a differential equation that presents some accuracy problem for very low values of the coupling constant and works better at higher values: the imaginary part is now  deduced from the half-width of the Breit-Wigner lines that fit the continuous spectrum. 
This makes  the two methods complementary and puts rather narrow bounds to the range of the ${\Omega}$ values in which the comparison makes sense: a reasonable interval for the present data could be assumed  to be $0.1\leq{\Omega}\leq0.25$.   Within these bounds Fig.1 shows that the agreement is complete,  proving the effectiveness of both the DBS and the method of \cite{GS} for solving numerical problems.


\section*{APPENDIX A.}
\label{appendixa}

\renewcommand{\theequation}{A.\arabic{equation}}
\setcounter{equation}{0}

In this Appendix we give the proof of the quadratic inequality (\ref{closure_cnd})
with $T_0$ and $T_1$ given in (\ref{H0H1_1dim}). We have:
\begin{eqnarray}
&{}& \!\!\!\!\!(T_0+T_1)^{\!*\,}(T_0+T_1)=(p^2+y^2+{\mathrm{Re}}(\sigma) y)^2\spazio{0.8}\cr
&{}&\phantom{XX}+( i\,{\mathrm{Im}}(\sigma)y+\frac{\sigma^2}{4}y^4 )\,(- i\,{\mathrm{Im}}(\sigma)y+\frac{{\bar\sigma}^2}{4}y^4 )
\spazio{0.8}\cr
&{}&\phantom{XX}
+( p^2+y^2+{\mathrm{Re}}(\sigma) y)\,( i\,{\mathrm{Im}}(\sigma)y+\frac{\sigma^2}{4}y^4 )\spazio{0.3}\cr
&{}&\phantom{XX}+(- i\,{\mathrm{Im}}(\sigma)y+\frac{{\bar\sigma}^2}{4}y^4 )\,(   p^2+y^2+{\mathrm{Re}}(\sigma) y)
\spazio{1.8}\cr
&{}&\!\!\!\!\!=~\Bigl(1-\frac{|{\mathrm{Re}}(\sigma^2)|}{|\sigma|^2}\Bigr)\,
(p^2+y^2+{\mathrm{Re}}(\sigma) y)^2\spazio{0.8}\cr
&{}&\phantom{XX}+({\mathrm{Im}}(\sigma)y^2+\frac14{\mathrm{Im}}(\sigma){\mathrm{Im}}(\sigma^2)y^5+\frac{|\sigma|^4}{16}y^8)
\spazio{0.8}\cr
&{}&\phantom{XX}+\frac{|{\mathrm{Re}}(\sigma^2)|}{|\sigma|^2}\,\Bigl[(p^2+y^2+{\mathrm{Re}}(\sigma)y\pm\frac{|\sigma|^2}{4}y^4)^2\spazio{0.8}\cr
&{}&\phantom{XX}-\frac{|\sigma|^4}{16}y^8\Bigr]
+{\mathrm{Im}}(\sigma)p+{\mathrm{Im}}(\sigma^2)(py^3+y^3p)
\nonumber
\end{eqnarray}
For $0\!<\!a\!<\!1\!-\!|{\mathrm{Re}}(\sigma^2)|/|\sigma|^2$ and for a certain $R\!>\!0$ we then have
\begin{eqnarray}
&{}&\!\!\!\!\! (T_0+T_1)^{\!*\,}(T_0+T_1)\geq a\Bigl[(p^2+y^2+{\mathrm{Re}}(\sigma) y)^2\cr
&{}&\phantom{XX}+({\mathrm{Im}}(\sigma)y^2+\frac14{\mathrm{Im}}(\sigma){\mathrm{Im}}(\sigma^2)y^5+\frac{|\sigma|^4}{16}y^8)
\Bigr]
\cr
&{}&\phantom{XX}+\frac{|{\mathrm{Re}}(\sigma^2)|}{|\sigma|^2}\,\Bigl[ {\mathrm{Im}}(\sigma)y^2+\frac14{\mathrm{Im}}(\sigma){\mathrm{Im}}(\sigma^2)y^5\Bigr]\cr
&{}&\phantom{XX}+R\Bigl[ (p^2+y^2+{\mathrm{Re}}(\sigma) y)^2+\frac{|\sigma|^4}{16}y^8 \Bigr]
\spazio{1.0}\cr
&{}&\phantom{XX}+\frac{{\mathrm{Im}}(\sigma)}{2}(p\pm1)^2-\frac{{\mathrm{Im}}(\sigma)}{2}(p^2+1)\cr
&{}&\phantom{XX}+|{\mathrm{Im}}(\sigma^2)|(p\pm y^3)^2-|{\mathrm{Im}}(\sigma^2)|(p^2+ y^6)
\spazio{1.8}\cr
&{}& \!\!\!\!\! \geq
\,\,\,a\,(T_0^*\,T_0+T_1^*\,T_1)\!-\!b\!+\!R\Bigl[(p^2+y^2+{\mathrm{Re}}(\sigma) y)^2\spazio{0.8}\cr
&{}&\phantom{XX}-|{\mathrm{Im}}(\sigma^2)|p^2-\frac{{\mathrm{Im}}(\sigma)}{2}(p^2+1)+\frac b2\Bigr]\spazio{1.0}\cr
&{}&\phantom{XX}+\Bigl[\frac{|{\mathrm{Re}}(\sigma^2)|}{|\sigma|^2}\,\Bigl(\frac14{\mathrm{Im}}(\sigma){\mathrm{Im}}(\sigma^2)y^5+ {\mathrm{Im}}(\sigma)y^2\Bigr)
\spazio{1.0}\cr
&{}&\phantom{XX}+R\frac{|\sigma|^4}{16}y^8-|{\mathrm{Im}}(\sigma^2)|y^6  +\frac b2 \Bigr]
\label{quadratic_estimate_1dim}
\end{eqnarray}

In the last inequality we have neglected two positive terms and we have chosen a  number $b>0$ in such a way to
make positive the last two square brackets. 
In spite of the explicit calculation, one could also have argued that the linear term  $\sigma y$ is small in the sense of Kato with respect to the leading term of the perturbation $(\sigma^2/4)y^4\,$: consequently it gives a negligible contribution to the quadratic estimate which therefore reduces to that of the anharmonic oscillator. We have chosen this shortcut to deal with the quadratic estimates of the three-dimensional case.





\begin{thebibliography}{999}
\bibitem{Dy} F. Dyson, \textit{Phys. Rev.} {85}, 631 (1952).
In this paper Dyson made the simple observation that  power series converge in disks, so that, if the continuation was possible, equation (\ref{SecondOrderEquationOmega_g}) would admit bound states also for $g^2\!\!<\!0$, which
does not occur. This signifies that the perturbation expansion in $g$ has zero radius of convergence
\end{thebibliography}
\end{document}